\documentclass{article}
\usepackage{waspaa21,amsmath,amssymb, graphicx,url,times, pifont}
\usepackage{color}
\usepackage{float}
\usepackage{multicol}
\usepackage{makecell}
\usepackage{subcaption}
\usepackage{algpseudocode,algorithm,algorithmicx}
\usepackage{amsfonts}
\usepackage{mathtools}
\usepackage{multirow}
\usepackage{bm}
\usepackage{nicefrac}
\usepackage{tabularx}
\usepackage{booktabs}
\usepackage{xcolor}
\usepackage{amsthm}
\usepackage{extarrows}
\usepackage{balance}
\usepackage{subcaption}
\newcommand*\Let[2]{\State #1 $\gets$ #2}
\algrenewcommand\algorithmicrequire{\textbf{Precondition:}}
\algrenewcommand\algorithmicensure{\textbf{Postcondition:}}
\newcommand{\eref}[1]{(\ref{#1})}

\newcommand{\fref}[1]{Fig.~\ref{#1}}

\newcommand{\xmark}{\ding{55}}
\renewcommand{\arraystretch}{1.1}
\newcommand{\ra}[1]{\renewcommand{\arraystretch}{#1}}

\title{Auto-DSP: Learning to Optimize Acoustic Echo Cancellers}

\name{Jonah Casebeer$^{\sharp}$\thanks{Code \& demo: \textit{https://jmcasebeer.github.io/projects/auto-dsp/}} \qquad 
Nicholas J. Bryan$^{\flat}$ \qquad 
Paris Smaragdis$^{\sharp \flat}$}
\address{
$^\sharp$ University of Illinois at Urbana-Champaign, $^\flat$ Adobe Research\\
}

\begin{document}

\ninept
\maketitle

\begin{sloppy}

\begin{abstract}
Adaptive filtering algorithms are commonplace in signal processing and have wide-ranging applications from single-channel denoising to multi-channel acoustic echo cancellation and adaptive beamforming. Such algorithms typically operate via specialized online, iterative optimization methods and have achieved tremendous success, but require expert knowledge, are slow to develop, and are difficult to customize. In our work, we present a new method to automatically learn adaptive filtering update rules directly from data. To do so, we frame adaptive filtering as a differentiable operator and train a learned optimizer to output a gradient descent-based update rule from data via backpropagation through time. We demonstrate our general approach on an acoustic echo cancellation task (single-talk with noise) and show that we can learn high-performing adaptive filters for a variety of common linear and non-linear multidelayed block frequency domain filter architectures. We also find that our learned update rules exhibit fast convergence, can optimize in the presence of nonlinearities, and are robust to acoustic scene changes despite never encountering any during training.
\end{abstract}


%%%%%%%%%%%%%%%%%%%%%%%%%%%%%%%%%%%%%%%%%%%%%%%%%%%%%%%%%%%%%%%%
%%%%%%%%%%%%%%%%%%%%%%%%%%%%%%%%%%%%%%%%%%%%%%%%%%%%%%%%%%%%%%%%
\section{Introduction}
Adaptive filtering algorithms are ubiquitous and include single- and multi-channel denoising, dereverberation, echo cancellation, system identification, noise cancellation, feedback cancellation, and more. Such algorithms typically operate by applying an online, iterative optimization method, such as least mean square filtering~(LMS), normalized LMS~(NLMS), or recursive least-squares~(RLS), to solve an optimization problem over time (e.g. estimating a time-varying transfer function for echo cancellation)~\cite{Widrow1985, hansler2005acoustic, benesty2001advances, haykin2008adaptive, apolinario2009qrd}. The derivation and implementation of these methods requires careful attention, customization, expertise, and/or a laborious tuning process (e.g. tuning per hardware device) per application.

One of the most prevalent adaptive filtering applications is acoustic echo cancellation (AEC). In this case, an adaptive filter is used to remove echo within a telecommunication system. Customized AEC adaptive filters take many forms including algorithms based on sparsity \cite{gay1998efficient}, adaptive normalization \cite{duttweiler2000proportionate}, and adaptive learning-rates \cite{valin2007adjusting}, as well as data-driven approaches for selecting learning rates automatically~\cite{dabney2012adaptive, mahmood2012tuning} and based on a meta-step-size~\cite{sutton1992adapting, schraudolph1999local}. More recently, deep learning techniques have been used as AEC sub-components including learned residual echo suppressors~\cite{zhang2019deep, fazel2020cad, valin2021low}, double-talk detectors \cite{ma2020acoustic}, and nonlinear distortions blocks~\cite{birkett1995acoustic, rabaa1998acoustic, zhang2017recursive, halimeh2019neural, malek2016hammerstein}. These approaches, however, commonly do not use neural network modules that adapt at test time, do not have matching training and testing steps, and/or do not directly learn adaptive filter update rules end-to-end.

In the machine learning literature, there have been exciting developments in meta-learning, automatic machine learning, and \emph{learning how to learn} methods. Methods include using one neural network to control the weights of another~\cite{schmidhuber1992learning, bello2017neural, ha2016hypernetworks}, pre-training deep networks that quickly fine-tune~\cite{finn2017model}, and learning offline stochastic gradient descent update rules via neural networks~\cite{andrychowicz2016learning}. The latter work is most relevant, shows how offline \emph{learned} optimizers can outperform their hand-designed counterparts for certain neural network architectures, and inspired further work in optimizer architectures~\cite{wichrowska2017learned} and training~\cite{metz2019understanding, chen2020training}. We believe this work is significant and, while not previously explored, offer tremendous potential for the field of signal processing and adaptive filtering.

\begin{figure}[t]
    \centering
    \includegraphics[width=.97\linewidth]{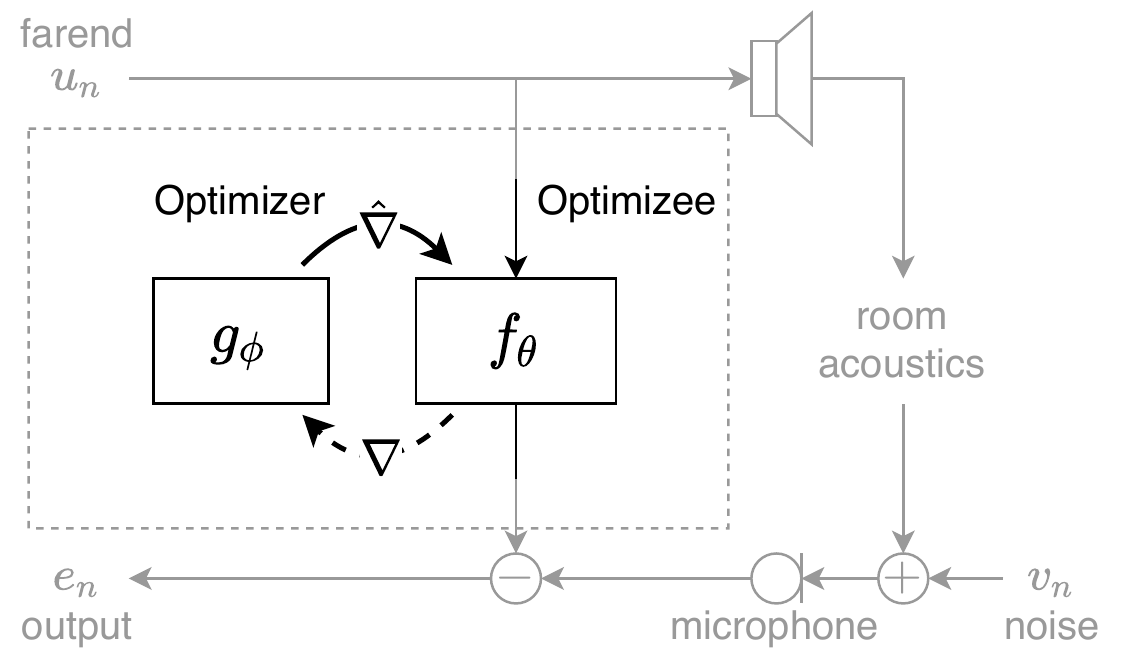}
    \vspace{-1.5mm}
    \caption{A learned optimizer, $g_\phi$, updates the adaptive filter $f_\theta$ in an online fashion. The optimizer parameters $\phi$ are meta-learned directly from data and do not use any external labels. The dashed curved line denotes adaptation during training, but not inference.}
    \label{fig:block_diagram}
    \vspace{-6mm}
\end{figure}

In this work, we formulate the development of adaptive filtering algorithms as a meta-learning problem and learn to optimize adaptive filters. To do so, we frame adaptive filtering itself as a differentiable operator and train a learned optimizer from data, without external labels, using truncated backpropagation through time. By doing so, we create an automatic digital signal processing~(Auto-DSP) approach that learns optimal adaptive filters without any need for hand-derived gradients and can be used for a variety of applications. To demonstrate our approach, we learn to optimize an AEC task as shown in~\fref{fig:block_diagram} for a single-talk in noise scenario. We use the Microsoft AEC Challenge dataset~\cite{sridhar2020icassp} to learn update rules for a variety of common linear and nonlinear multidelayed block frequency domain filters~(MDF)~\cite{soo1990multidelay}. We compare our results to hand-engineered, grid-search-tuned block NLMS and RMSprop~\cite{hinton2012neural} optimizers, as well as the open-source Speex AEC~\cite{valin2007adjusting, valin2016speex}. We find our learned optimizers outperform all of these methods for our tasks, exhibit fast convergence, require little-to-no manual design intervention for training, can optimize in the presence of nonlinearities, and converge quickly and robustly to unseen acoustic scene changes.

%%%%%%%%%%%%%%%%%%%%%%%%%%%%%%%%%%%%%%%%%%%%%%%%%%%%%%%%%%%%%%%%
%%%%%%%%%%%%%%%%%%%%%%%%%%%%%%%%%%%%%%%%%%%%%%%%%%%%%%%%%%%%%%%%
\section{Auto-DSP Optimization}
To learn an adaptive filter update rule from data,
%we modify past offline learning to learn techniques~\cite{andrychowicz2016learning} to the adaptive filtering signal processing domain. First,
we first define a learned optimizer, $g_\phi(\cdot)$, as a function doing the optimizing and an optimizee, $f_\theta(\cdot)$, as a differentiable adaptive filter to be optimized, and $J(\cdot)$ as the optimizee loss. Second, we set the optimizer to be a neural network that accepts as input raw optimizee gradients $\nabla f_{\theta_n}(\cdot)$ and a state vector $\mathbf{h}$ and outputs a learned gradient descent update rule,
\begin{equation}
\theta_{n+1} = \theta_{n} + g_\phi(\nabla f_{\theta_n}(\cdot), \mathbf{h}),
\label{eq:update}
\end{equation}
where $\theta$ are optimizee parameters and $\phi$ are optimizer parameters. The state, $\mathbf{h}$, is also updated. Note, the raw gradient inputs used here are provided by automatic differentiation and are not implemented manually.
Third, we assign the optimizer and optimizee an objective function or loss and use truncated backpropagation through time (BPTT)~\cite{werbos1990backpropagation} to fit the optimizer parameters to data. %Details on our optimizee, optimizer, and training algorithm are found below.

%%%%%%%%%%%%%%%%%%%%%%%%%%%%%%%%%%%%%%%%%%%%%%%%%%%%%%%%%%%%%%%%%%%%%%%%%%%%%%%%
\subsection{Optimizee architecture \& loss}
The optimizee, or adaptive filtering being optimized, provides the architecture used for filtering signals. It is defined by filter parameters $\theta$, a filtering architecture $f_\theta(\cdot)$, and an optimizee loss function. For illustrative purposes, we can consider a basic time-domain adaptive filter optimizee. In this case, the optimizee parameters $\theta$ correspond to transversal finite impulse response (FIR) filter coefficients $\theta = \{\hat{\mathbf{w}}_n \in \mathbb{R}^N\}$, the optimizee architecture corresponds to the inner product between an input vector $\mathbf{u}_n \in \mathbb{R}^N$ and the filter coefficients $f_\theta(\mathbf{u}_n) = y_n = \hat{\mathbf{w}}_n^{\mathsf{H}} \mathbf{u}_n$, and the optimizee loss corresponds to a mean squared error objective, $J(y_n, d_n) = \frac{1}{N}\sum_n^N |y_n - d_n|^2$, where $d_n \in \mathbb{R}$ is the desired, known response.
In this case, we can reduce the optimizee update~\eref{eq:update} to
\begin{equation}
    \hat{\mathbf{w}}_{n+1} = \hat{\mathbf{w}}_n  - g_{\phi}( \mathbf{u}_n \cdot (\hat{\mathbf{w}}_n^{\mathsf{H}} \mathbf{u}_n - d_n)^*),
    \label{eq:lms_optimizee}
\end{equation}
where $^*$ denotes complex conjugation and $^{\mathsf{H}}$ denotes Hermitian transposition.

While we manually derive the gradient vector $\nabla f_{\theta_n}(\cdot)$ here, in practice gradients are computed via automatic differentiation.
Because of this, we can use more advanced optimizees such as lattice FIR filters, block frequency-domain filters~\cite{Widrow1985, haykin2008adaptive}, multidelayed block frequency domain filters~\cite{soo1990multidelay}, or non-linear variants such as (polynomial) Volterra filters~\cite{mathews1991adaptive}, or Hammerstein filters~\cite{scarpiniti2014hammerstein} with ease. We can also use alternative differentiable optimizee losses such as negative log-likelihood or mutual information. %In this way, we can easily prototype and learn adaptive algorithms in an automatic way.

%%%%%%%%%%%%%%%%%%%%%%%%%%%%%%%%%%%%%%%%%%%%%%%%%%%%%%%%%%%%%%%%%%%%%%%%%%%%%%%%
\subsection{Optimizer architecture \& loss}
The optimizer $g_\phi(\cdot)$, or function doing the optimizing, is parameterized by $\phi$ and used to adapt the optimizee parameters $\theta$ over time. The optimizer accepts as input raw optimizee gradients and outputs an optimized, learned update rule. For a basic time-domain adaptive filter optimizee, we can define the optimizer architecture as a single step-size $\mu$ and reduce~\eref{eq:lms_optimizee} to
\begin{equation}
    \hat{\mathbf{w}}_{n+1} = \hat{\mathbf{w}}_n  - \mu \cdot \mathbf{u}_n(\hat{\mathbf{w}}_n^{\mathsf{H}} \mathbf{u}_n - d_n)^*,
\end{equation}
or the well known LMS algorithm.
For a more powerful optimizer, however, we can define the optimizer $g_{\phi}(\nabla J(\cdot))$ to be a neural network module such as a recurrent neural network (RNN), convolutional neural network (CNN), fully connected network, or similar. The design of the optimizer, however, has tremendous implications on computational complexity of the approach.
Thus, we make the optimizer agnostic of the optimizee layout by applying the optimizer independently to each element of the optimizee parameters $\theta$. That is, the optimizer update is applied element-wise to each optimizee parameter. This allows us to efficiently vectorize our update rules and perform weight sharing in the optimizer, while maintaining independent state dynamics per optimizee parameter.

In terms of the optimizer objective, we set it to be the sum of a collection of optimizee losses averaged across a dataset, which requires no additional labels. However, this can be modified to favor different optimization dynamics or design constraints. For example, we could use a weighted sum where earlier (or later) losses are weighted more. This would enable training an optimizer specializing in early (or late) convergence. Other schemes could produce optimizers that favor small updates, sparse updates, etc.

%%%%%%%%%%%%%%%%%%%%%%%%%%%%%%%%%%%%%%%%%%%%%%%%%%%%%%%%%%%%%%%%%%%%%%%%%%%%%%%%
\subsection{Learning the optimizer}
To train our optimizers, we follow the procedure outlined in Algorithm \ref{alg:inner_outer}. The procedure consists of two basic steps: an inner loop and an outer loop. The inner loop process runs an adaptive filter optimizee for a finite number of time steps, $N_i$, updates the optimizee parameters as it goes, and accumulates the optimizee loss for a fixed optimizer state using BPTT. The outer loop invokes the inner loop, uses a standard deep learning optimizer denoted as a meta optimizer (\textproc{MetaOpt}) to update the learned optimizer module, and repeats for a finite number of outer loop steps, $N_o$. In practice, the outer loop is vectorized and runs across a randomized collection of signals (i.e. batches) continuously sampled from data until the optimizer loss convergences. This procedure allows us to train our optimizer on long sequences and helps minimize exploding gradient issues. After training, the learned optimizers are used like conventional optimizers and do not use the inner/outer scheme.

\begin{algorithm}[t]
  \caption{Meta-learning training algorithm.}
  \begin{algorithmic}[1]
    \Require{$J(\cdot)$, \textproc{MetaOpt}, $N_i$, $N_o$}
    \Function{InnerLoop}{$\phi, \mathbf{h}, \theta, \mathbf{u}, \mathbf{d}$}
      \Let{$\mathcal{L}$}{0}
      \For{$n_i \gets 0 \textrm{ to } N_i$}
          \Let{$y_{n_i}$}{$f_{\theta_{n_i}}(\mathbf{u}_{n_i})$}
          \Let{$\mathcal{L}$}{$\mathcal{L} +  J(y_{n_i}, \mathbf{d}_{n_i})$}
          \Let{$\hat{\nabla}, \mathbf{h}$} {$g_{\phi}(\nabla J(y_{n_i}, \mathbf{d}_{n_i}), \mathbf{h})$}
          \Let{$\theta_{n_i + 1}$}{$\theta_{n_i} - \hat{\nabla}$}
      \EndFor
      \State \Return{$\mathcal{L}, \theta_{N_i}, \mathbf{h}$}
    \EndFunction
    \Function{OuterLoop}{$\phi, \mathbf{h}, \theta, \mathbf{u}, \mathbf{d}$}
      \For{$n_o \gets 0 \textrm{ to } N_o$}
          \Let{$\mathcal{L}, \theta_{n_o + 1}, \mathbf{h}$}{\textproc{InnerLoop}($\phi_{n_o}, \mathbf{h}, \theta_{n_o}, \mathbf{u}_{n_o}, \mathbf{d}_{n_o}$)}
          \Let{$\phi_{n_o + 1}$}{\textproc{MetaOpt}($\phi_{n_o}, \mathcal{L}$)}
      \EndFor
      \State \Return{$\phi_{N_o}$}
    \EndFunction
  \end{algorithmic}
  \label{alg:inner_outer}
\end{algorithm}

%%%%%%%%%%%%%%%%%%%%%%%%%%%%%%%%%%%%%%%%%%%%%%%%%%%%%%%%%%%%%%%%%%%%%%%%%%%%%%%%

%%%%%%%%%%%%%%%%%%%%%%%%%%%%%%%%%%%%%%%%%%%%%%%%%%%%%%%%%%%%%%%%
%%%%%%%%%%%%%%%%%%%%%%%%%%%%%%%%%%%%%%%%%%%%%%%%%%%%%%%%%%%%%%%%
\section{Experimental Setup}

\subsection{Optimizee configuration}

To demonstrate our approach, we consider the adaptive filtering task of acoustic echo cancellation or interference cancellation. For our AEC optimizee architecture, we use an MDF filter with an optional parametric nonlinearity. The optimizee parameters $\theta$ include frequency domain filter coefficients and a small set of nonlinear coefficients. The filter coefficients are partitioned into multiple delayed blocks and used within the framework of overlap-save short-time Fourier transform processing~\cite{rabiner2016theory}. MDF filters are commonly used for AEC and leverage the benefits of both frequency-domain adaptation~\cite{haykin2008adaptive} and low latency. For our optimizee loss, which implicitly defines the optimizer loss, we use the mean squared error.

In more detail, our MDF filter consists of frequency domain filter coefficients $\mathbf{W} \in \mathbb{C}^{M \times N}$, where $M$ is the number of delayed blocks, $N$ is the fast Fourier transform (FFT) size, $P=M\cdot N/2$ is the number of filter parameters, and $L$ is the filter length in samples. The filter matrix is applied to the delayed frequency domain near-end inputs $\mathbf{U} \in \mathbb{C}^{M \times N}$ to yield a filtered output via $y_n = \text{last } N/2 \text{ terms of} \{ \operatorname{FFT}^{-1}((\mathbf{W} \odot \mathbf{U})^\top \mathbf{1}_N) \}$, where $^\top$ is a matrix transpose, $\odot$ is the hadamard product, and $\mathbf{1}_N$ is an $N \times 1$ matrix of ones. To construct $\mathbf{U}$, we buffer the time-domain near-end signal to length $N$ with time overlap $R$, forming $\mathbf{u}_{\tilde{n}} \in \mathbb{R}^N$, shift $\mathbf{U}_m = \mathbf{U}_{m+1}$ for $m = 1, 2, \cdots, M -1$, and assign $\mathbf{U}_M = \operatorname{FFT}(\mathbf{u}_{\tilde{n}})$. Finally, we antialias $\mathbf{W}$ after each update so that each block has $N/2$ nonzero time-domain parameters. For our nonlinearity extension, we preprocess each element $u_n$ of the far-end reference signal through a parametric sigmoid
\begin{align}
    \gamma(u_n) &= \alpha_4 \left(\frac{2}{1 + \exp(\alpha_2 \hat{u}_n  + \alpha_3 \hat{u}_n^2)} - 1\right)
\end{align}
where $\hat{u}_n = (u_n \cdot \alpha_1)/(\sqrt{|u_n|^2 + |\alpha_1|^2})$ and $\alpha_i \forall i$ are adapted.

%%%%%%%%%%%%%%%%%%%%%%%%%%%%%%%%%%%%%%%%%%%%%%%%%%%%%%%%%%%%%%%%
\subsection{Optimizer configuration}
We implement a complex-valued gated recurrent unit~(GRU) optimizer architecture composed of a complex linear layer with output size $H$, two weight-tied complex GRU layers, and a complex output linear layer. The GRU layers share a single $H$ dimensional hidden state $\mathbf{h} \in \mathbb{C}^{H}$. All layers are followed by complex rectified linear units (ReLU). The GRU layers use complex-valued hyperbolic tangent (tanh) and sigmoid activation functions. We use the initialization scheme proposed in~\cite{wolter2018complex} and test two optimizer sizes: $H=24$ and $H=48$ with 3.6k and 14k complex-valued parameters, respectively. To train this model, we use the mean-squared error optimizee loss averaged across a batch of optimizees. %A single forward pass when $H=48$ and the optimizee has $P=2048$ parameters requires $\approx 240$ mega-MACS.

As a preprocessing step to our GRU-based optimizer, we modify the feature extraction in~\cite{andrychowicz2016learning} to operate in the complex domain. Specifically, we input complex optimizee gradients $\nabla$ and then limit the dynamic range by clipping and compressing the gradient magnitudes via:
\begin{equation}
    \tilde{\nabla} = \frac{\log(\max(e^{-p}, \min(|\nabla|,e^p ))) + p}{p} e^{j \angle \nabla},
\end{equation}
where $p$ is a hyperparameter that controls the clipping and $e$ is the exponential function. The purpose of this is to leave the phase of the gradient unchanged. We set $p=10$ in all experiments.
%%%%%%%%%%%%%%%%%%%%%%%%%%%%%%%%%%%%%%%%%%%%%%%%%%%%%%%%%%%%%%%%
\subsection{Evaluation metrics}
To measure the average performance of our learned adaptive filters and compare to known baselines, we use the average echo return loss enhancement (ERLE), which is defined as $10 \log({\sum_n |d_n|^2}/{\sum_n |y_n-d_n|^2})$. To understand convergence speed and provide further qualitative analysis, we also use the segmental ERLE, or the ERLE computed on short windows of size $N$.

%%%%%%%%%%%%%%%%%%%%%%%%%%%%%%%%%%%%%%%%%%%%%%%%%%%%%%%%%%%%%%%%
\subsection{Dataset}
We train and test our optimizers using the synthetic portion of the Microsoft AEC Challenge dataset~\cite{sridhar2020icassp}. This dataset includes far-end noise, near-end noise, and far-end nonlinearities. We preprocess the data by resampling to $8$kHz and remove near-end speech from all near-end recordings for focus and leave learning double-talk robust optimizers for future work. Note that the dataset is composed of $80\%$ nonlinear scenes and $20\%$ linear scenes.

%%%%%%%%%%%%%%%%%%%%%%%%%%%%%%%%%%%%%%%%%%%%%%%%%%%%%%%%%%%%%%%%
\subsection{Training}
For training, we follow Algorithm~\ref{alg:inner_outer} together with a standard training, validation, and testing setup. We use Adam~(lr$=10^{-4}$) with gradient clipping as our meta-optimizer, and halve the learning rate if ERLE performance on the validation set does not improve for $10$ consecutive epochs and cease training after $25$ epochs with no ERLE improvement. We define an epoch to be $200$ batches of optimization runs where each optimization run consists of an initial filter state, a $10$ second far-end signal and a $10$ second near-end signal. We alternate between inner and outer updates as defined in Algorithm \ref{alg:inner_outer} and set $N_i=10$. $N_o$ is set to be the number of inner loop updates that will fit within a $10$ second sequence. We found that optimizers did not converge well when the value of $N_i$ was set much higher than $20$ or lower than $5$. Our implementation is written using the JAX framework~\cite{jax2018github}. Training an optimizer takes one to ten days depending on optimizer/optimizee complexity on two RTX 2080 TI GPUs.

%%%%%%%%%%%%%%%%%%%%%%%%%%%%%%%%%%%%%%%%%%%%%%%%%%%%%%%%%%%%%%%%

\section{Results \& Discussion}
We evaluate our learned optimizers across multiple optimizer configurations and optimizee configurations as well as linear and nonlinear scenes and compare against standard hand-derived update rules.  Our baselines consist of step-size tuned block frequency-domain NLMS optimizer with smoothing constant~($\beta=.9, .99$), a step-size tuned frequency-domain RMSprop optimizer, and the Speex AEC. While Speex is representative of a well-engineered hand-tuned optimizer it was not optimized for this dataset whereas the other optimizers are. Baseline results are shown in the first section of Table~\ref{table:all_results}. We denote the learned optimizer hidden size by $H$, the number of filter parameters as $P$, the FFT size as $N$, the number of MDF blocks as $M$, the overlap between blocks as $R$, whether the optimizee has a nonlinear component with $\gamma$, and provide both the average $\mu$ and standard deviation $\sigma$ ERLE. All baseline and learned optimizees have an effective filter length of $L=2048$ taps.

%%%%%%%%%%%%%%%%%%%%%%%%%%%%%%%%%%%%%%%%%%%%%%%%%%%%%%%%%%%%%%%%
\subsection{Feature extraction}
We evaluate our proposed feature extraction by training optimizers ($H=24, P=2048, N=2P, M=1, R=1/2$, and no $\gamma$) with and without this preprocessing and display the results in the second portion of Table~\ref{table:all_results}. As shown, our proposed features improve performance by $\approx2$dB and $\approx3$dB dB in nonlinear and linear scenes, respectively. By inspection, we found that that the distribution of gradient magnitudes was heavily skewed and we hypothesize that clipping and compressing gradient magnitudes acts as a form of whitening that approximately normalizes the distribution. Given this result, we use the proposed feature extraction for all further experiments.

%%%%%%%%%%%%%%%%%%%%%%%%%%%%%%%%%%%%%%%%%%%%%%%%%%%%%%%%%%%%%%%%
\subsection{Optimizer capacity \& computational complexity}
\label{sec:capacity}
Next, we increase the hidden state size $H$ to $48$ and compare different proportions of overlap in the third portion of Table~\ref{table:all_results}. At $R=1/2$ and $3/4$, the learned optimizer outperforms NLMS and RMSprop. When $R$ is increased to $7/8$, performance improves by multiple dB and the learned optimizer outperforms all baselines with a slightly higher standard deviation than Speex. For this configuration, the optimizee has $P=2048$ and uses $240$ mega-MACS per update. On an i9-9820X CPU the real-time factor is $.36$ using one thread and $.13$ using multiple threads. This is remarkable, given the engineering expertise and effort distilled into our baselines.

%%%%%%%%%%%%%%%%%%%%%%%%%%%%%%%%%%%%%%%%%%%%%%%%%%%%%%%%%%%%%%%%
\subsection{Learned optimizer dynamics}

\setlength{\tabcolsep}{1.75pt}
\begin{table}[t]
    \centering
    \ra{.90}
    \begin{tabular}{@{}l c c c c c c c c c c@{}}\toprule
        Optimizer & \multicolumn{4}{c}{Optimizee} &&  \multicolumn{2}{c}{\thead{ERLE~(dB)\\Nonlinear}} && \multicolumn{2}{c}{\thead{ERLE~(dB)\\Linear}} \vspace{-2mm}\\
        \cmidrule{2-5} \cmidrule{7-8} \cmidrule{10-11} & $M$ & $N$ & $R$ & $\gamma$ && $\mu$ & $\sigma$ && $\mu$ & $\sigma$\vspace{-.5mm}\\
        \midrule
        NLMS~($\beta=.9$) & $1$ & $4096$ & $7/8$ & \xmark&& 4.40 & 11.82 && 9.57 & 6.01\\
        NLMS~($\beta=.99$) & $1$ & $4096$  & $7/8$ & \xmark&& 4.07 & 4.54 && 5.20 & 2.57\\
        RMSprop & $1$  & $4096$ & $7/8$ & \xmark && 5.58 & 2.94 && 7.71 & 2.56\\
        Speex & $4$  & $1024$ & $1/2$ & \xmark && 9.79 & 4.56 && 8.55 & 3.51\\
        Speex & $8$  & $512$ & $1/2$ & \xmark && 9.83 & 4.50 && 8.72 & 3.55 \vspace{-.5mm} \\
        \midrule
        GRU~($H=24, \nabla$) & $1$  & $4096$ & $1/2$ & \xmark && 3.24 & 1.78 && 4.68 & 1.85\\
        GRU~($H=24, \tilde{\nabla}$)   & 1 & $4096$ & $1/2$ & \xmark && 5.69 & 2.86 && 7.78 & 2.13 \vspace{-.5mm}\\
        \midrule
        GRU~($H=48, \tilde{\nabla}$) & $1$ & $4096$ & $1/2$ & \xmark && 5.87 & 2.93 && 7.95 & 2.15\\
        GRU~($H=48, \tilde{\nabla}$) & $1$ & $4096$ & $3/4$ & \xmark && 8.26 & 4.03 && 11.22 & 2.93\\
        GRU~($H=48, \tilde{\nabla}$) & $1$ & $4096$ & $7/8$ & \xmark && 10.40 & 5.18 && 14.21 & 4.29 \vspace{-.5mm}\\
        \midrule
        GRU~($H=48, \tilde{\nabla}$) & $4$  & $1024$ & $1/2$ & \xmark && 8.11 & 4.40 && 9.46 & 3.26\\
        GRU~($H=48, \tilde{\nabla}$) & $4$  & $1640$ & $3/4$ & \xmark && 10.20 & 5.15 && 13.62 & 3.87\\
        GRU~($H=48, \tilde{\nabla}$) & $8$  & $512$ & $1/2$ & \xmark && 8.45 & 4.58 && 9.54 & 3.32\\
        GRU~($H=48, \tilde{\nabla}$) & $8$  & $912$ & $3/4$ & \xmark && 10.75 & 5.46 && 13.93 & 3.94 \vspace{-.5mm}\\
        \midrule
        GRU~($H=48, \tilde{\nabla}$) & $1$ & $4096$ & $7/8$ & \checkmark && 10.53 & 4.04 && 13.45 & 3.55\\
        GRU~($H=48, \tilde{\nabla}$) & $4$ & $1640$ & $3/4$ & \checkmark && 9.17 & 3.73 && 11.66 & 4.02\\
        GRU~($H=48, \tilde{\nabla}$) & $8$ & $912$ & $3/4$ & \checkmark && 10.17 & 4.12 && 11.61 & 4.07 \vspace{-.5mm}\\
        \bottomrule
        \end{tabular}
    \vspace{-1.5mm}
    \caption{Optimizer comparison using the ERLE metric. The optimizee column shows the number of blocks $M$, the FFT size $N$, the proportion of block overlap $R$, and the optimizee nonlinearity extension $\gamma$. All optimizees have a filter length of $L=2048$.}
    \label{table:all_results}
    \vspace{-4.5mm}
\end{table}

We evaluate how our learned optimizers respond to an abrupt change in the echo path using the final setup from Section~\ref{sec:capacity}. The learned optimizers were trained on static scenes with a duration of $10$ seconds. However, test scenes here are twice as long and formed by concatenating two test set files. The Block NLMS and RMSprop baselines were tuned on static scenes to match the learned optimizer setup. In Fig.~\ref{fig:fb_scene_change}, we compare ERLE  across time. To match the number of updates, we run Speex with $4$ blocks.

In both linear and nonlinear scenes the learned optimizer converges rapidly and achieves a steady-state in $\approx 2$ seconds. In linear scenes, block NLMS is competitive and reaches a steady-state ERLE at $\approx 6$ seconds. However, after the scene change at $10$ seconds, the learned optimizer is the only optimizer that recovers its full performance. In nonlinear scenes, Speex displays strong steady-state performance and overtakes the learned optimizer. Though, after the scene change, the learned optimizer recovers and is not surpassed. On average, our optimizer outperforms all baselines for both steady-state and early convergence and demonstrates it can generalize to novel and challenging environments.

%%%%%%%%%%%%%%%%%%%%%%%%%%%%%%%%%%%%%%%%%%%%%%%%%%%%%%%%%%%%%%%%
\subsection{Optimizee architecture}
In our final set of experiments, we learn optimizers for a variety of optimizee architectures. That is, we construct MDF adaptive filters with $4$ and $8$ blocks (instead of one block) and also experiment with incorporating a parametric nonlinear distortion block and adjusting the overlap $R$. Results can be found in the last two sections of Table~\ref{table:all_results}. Note that these optimizees may have $P>L$ parameters. For all adaptive filters, we use identical optimizer hyperparameters, training scheme, and architecture. In effect, no manual architecture design intervention is required for any of our learned optimizers.

First, we find that our learned optimizers successfully scale to more complex linear adaptive filter architectures. Second, we find that we can learn to optimize nonlinear variants of MDF, and generally outperform their hand-tuned counterparts in nonlinear scenes.
This suggests we can learn optimal update rules per filtering architecture to fit design trade-offs like latency versus computational complexity without needing to hand-derive anything and has the potential to fundamentally change how we develop adaptive filters.

\begin{figure}[t]
    \centering
    \includegraphics[width=\linewidth]{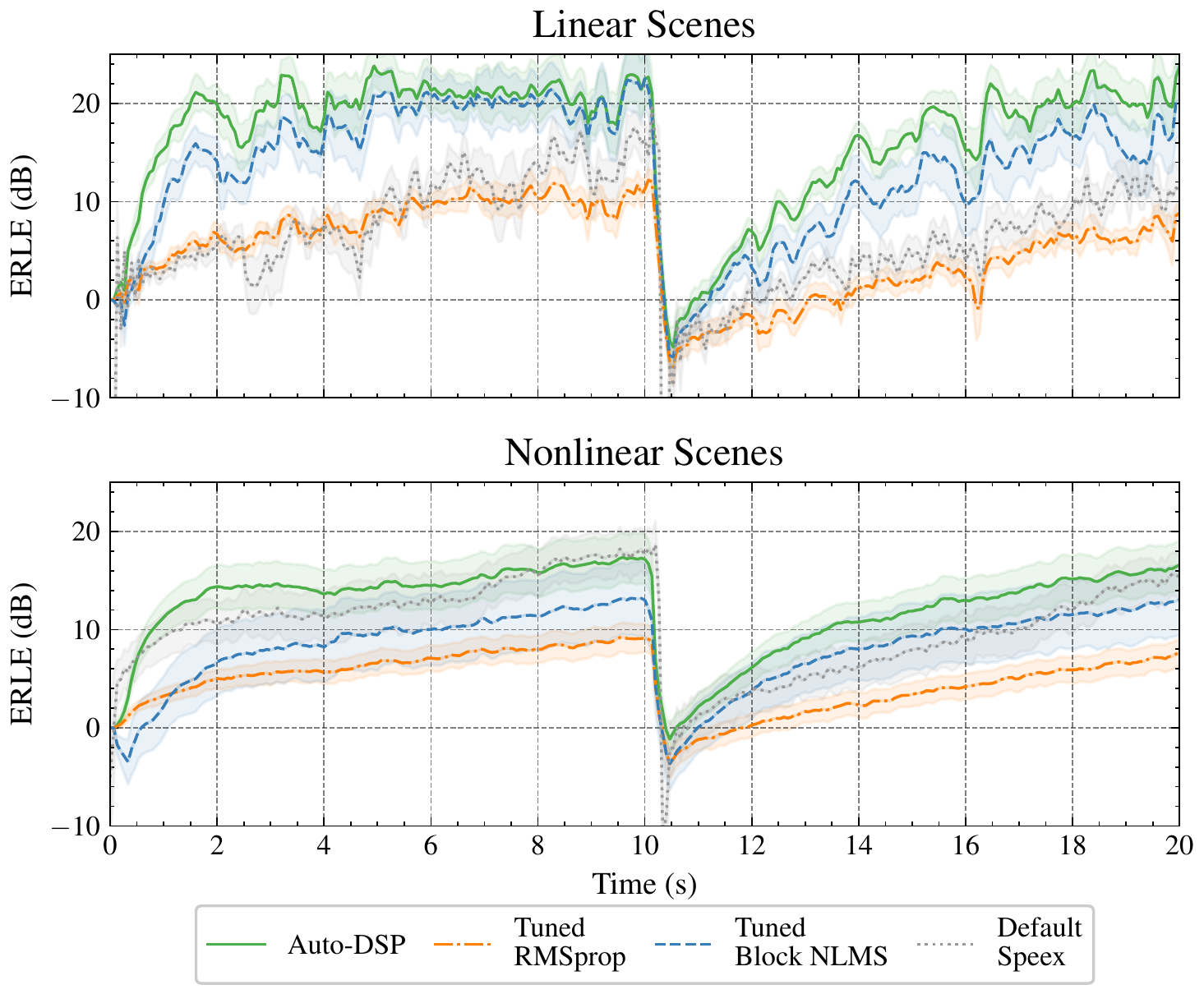}
    \vspace{-4mm}
    \caption{ERLE~(dB) performance across time in linear~(top) and nonlinear~(bottom) scenes. The echo path changes ten seconds into the scene. Lines represent mean performance and shaded regions represent $\pm \frac{1}{2}$ a standard deviation. Our optimizer converges faster than other optimizers and quickly adapts despite being trained on shorter scenes and never encountering scene changes in training.}
    \label{fig:fb_scene_change}
    \vspace{-4mm}
 \end{figure}

\section{Conclusion}
In this work, we formulate the optimization of adaptive filters as a meta-learning problem and successfully replace hand-derived update rules with learned update rules. We call this method Auto-DSP and apply it in acoustic echo cancellation where we learn optimization rules in a data-driven fashion, without any external labels. Using an identical optimizer configuration, we experiment with learning update rules for multidelay block frequency domain filters both with and without parametric nonlinearities. We evaluate performance across scenes with nearend noise and far-end distortion, and find we can outperform tuned block NLMS and RMSprop optimizers and a popular open source filter (Speex). In all, we believe learning adaptive filter update rules from data is an exciting new signal processing methodology and has tremendous potential.

% -------------------------------------------------------------------------
% Either list references using the bibliography style file IEEEtran.bst
\bibliographystyle{IEEEtran}
\bibliography{refs21}
%
% or list them by yourself
% \begin{thebibliography}{9}
% 
% \bibitem{waspaa21web}
%   \url{http://www.waspaa.com}.
%
% \bibitem{IEEEPDFSpec}
%   {PDF} specification for {IEEE} {X}plore$^{\textregistered}$,
%   \url{http://www.ieee.org/portal/cms_docs/pubs/confstandards/pdfs/IEEE-PDF-SpecV401.pdf}.
%
% \bibitem{PDFOpenSourceTools}
%   Creating high resolution {PDF} files for book production with 
%   open source tools, 
%   \url{http://www.grassbook.org/neteler/highres_pdf.html}.
%
% \bibitem{eWilliams1999}
% E. Williams, \emph{Fourier Acoustics: Sound Radiation and Nearfield Acoustic
%   Holography}. London, UK: Academic Press, 1999.
% 
% \bibitem{ieeecopyright}
%   \url{http://www.ieee.org/web/publications/rights/copyrightmain.html}.
%
% \bibitem{cJones2003}
% C. Jones, A. Smith, and E. Roberts, ``A sample paper in conference
%   proceedings,'' in \emph{Proc. IEEE ICASSP}, vol. II, 2003, pp. 803--806.
% 
% \bibitem{aSmith2000}
% A. Smith, C. Jones, and E. Roberts, ``A sample paper in journals,'' 
%   \emph{IEEE Trans. Signal Process.}, vol. 62, pp. 291--294, Jan. 2000.
% 
% \end{thebibliography}

\end{sloppy}
\end{document}